\documentclass[runningheads]{llncs}
\usepackage[T1]{fontenc}
\usepackage{graphicx}
\usepackage{hyperref}
\usepackage{comment}
\usepackage{xcolor}
\usepackage{color}

\usepackage{amsmath}
\usepackage{amsfonts}
\usepackage{nicefrac}

\newcommand\extrafootertext[1]{%
    \bgroup
    \renewcommand\thefootnote{\fnsymbol{footnote}}%
    \renewcommand\thempfootnote{\fnsymbol{mpfootnote}}%
    \footnotetext[0]{#1}%
    \egroup
}

\usepackage[labelformat=simple]{subcaption}

\usepackage[numbers,sort&compress, sectionbib]{natbib}

\usepackage{acronyms}

\begin{document}
\sloppy
%
\title{Short Paper: Atomic Execution is Not Enough for Arbitrage Profit Extraction in Shared Sequencers}
\titlerunning{ }
%
\author{Maria Inês Silva\inst{1, 3} \and Benjamin Livshits\inst{2}}
\authorrunning{ }
\institute{Imperial College London \and NOVA Information Management School \and Matter Labs}
\extrafootertext{$^*$ Some of this work was performed while the authors were at Matter Labs.}
\maketitle              

\begin{abstract}

\vspace{-3ex}
There has been a growing interest in shared sequencing solutions, in which transactions for multiple rollups are processed together. Their proponents argue that these solutions allow for better composability and can potentially increase sequencer revenue by enhancing MEV extraction. However, little research has been done on these claims, raising the question of understanding the actual impact of shared sequencing on arbitrage profits, the most common MEV strategy in rollups. To address this,  we develop a model to assess arbitrage profits under atomic execution across two Constant Product Market Marker liquidity pools and demonstrate that switching to atomic execution does not always improve profits. We also discuss some scenarios where atomicity may lead to losses, offering insights into why atomic execution may not be enough to convince arbitrageurs and rollups to adopt shared sequencing.

\end{abstract}

\section{Introduction}
\label{sec:intro}

\gls{DeFi} has been essential to the growth of the Ethereum ecosystem, attracting many users and successful applications. Recently, it has expanded to \gls{L2} scaling solutions like rollups, where trading volumes are rising, with some rollups now experiencing more daily activity than Ethereum itself~\cite{gogol_layer-2_2024}. With this growth in adoption comes more opportunities for \gls{MEV}~---~a collection of techniques for extracting value from transaction inclusion and reordering~\cite{daian_flash_2020}. One of the most prevalent forms of \gls{MEV} on rollups is arbitrage, in which arbitrageurs exploit price differences between centralized exchanges and/or \glspl{DEX}~\cite{torres_rolling_2024, oz_pandoras_2025}.

A relevant consideration for \gls{MEV} in rollups is \emph{sequencer design}. The sequencer is the operator responsible for receiving and scheduling user transactions for processing, and most rollups currently use an independent centralized sequencer~\cite{motepalli_sok_2023}. Recent proposals have introduced an alternative~---~shared sequencers. Shared sequencing schemes propose to process transactions for multiple rollups together, allowing for better composability between rollups. Despite being a recent topic, Astria~\cite{noauthor_astria_2024} already has a solution in production, while Radius~\cite{radius_radius_2023}, NodeKit~\cite{nodekit_composable_2024}, and Expresso Systems~\cite{espresso_systems_espresso_2023} are in the test phase. However, we have not yet seen significant adoption from rollups.

Proponents of shared sequencing argue that it can enhance \gls{MEV} extraction in cross-rollup arbitrage, thus adding a potential for increased revenue for rollups. Arbitrageurs can already execute cross-rollup arbitrage by submitting independent transactions to each rollup. However, this strategy involves additional liquidity and currency risk costs.

In this context, shared sequencing offers two relevant properties for arbitrageurs: \textit{atomic execution} and \textit{atomic bridging}. Atomic execution allows an arbitrageur to bundle two swaps (one for each rollup) and have the guarantee that if one of the swaps reverts, the other will also revert. This property requires control over block-building on the rollups running full nodes of the rollups to guarantee execution validity. Atomic bridging goes further by allowing for bridge operations between rollups, eliminating the need for liquidity across different chains. An arbitrageur can take a flash loan on one rollup, swap tokens, bridge them to another rollup for the second swap, and then bridge the tokens back to repay the loan. Yet, this property is significantly more challenging to achieve and requires additional trust assumptions on the shared sequencing infrastructure.

With the added complexities of implementing atomic bridging, in this work, we aim to understand how arbitrage profits can be impacted by a shared sequencing solution that only provides atomic execution. Even though this property seems intuitively beneficial for arbitrageurs, we argue this is not always true. In fact, atomic execution is insufficient to consistently improve \gls{MEV} extraction for arbitrageurs and therefore to increase revenue for sequencers.

Concretely, we build a model to assess the difference in terms of the expected arbitrage profit of switching to a shared sequencing regime with atomic execution. Here, we consider a cross-rollup arbitrage between two \gls{CPMM} liquidity pools and compute the expected profit obtained by the arbitrageur given key parameters such as the prices in the pools and the probabilities of failure of the swaps. Then, we analyze how this difference in expected profit changes with these parameters and conclude that an arbitrageur does not always benefit from atomic execution. We also discuss and provide some intuition as to why atomic execution leads to losses in some particular cases.

These results are consistent with previous work from~\citet{mamageishvili_shared_2023}, in which they consider how atomic execution impacts arbitrageurs' latency competition and their incentives to invest in latency. Interestingly, they observe that in a regime where transaction order and inclusion is determined through bidding, the revenue of shared sequencing is not always higher than that of separate sequencing and depends on the transaction ordering rule applied and the arbitrage value potentially realized.

\section{Modelling Arbitrage Extraction}
\label{sec:model}

Before describing the model to estimate the impact of atomic execution for cross-rollup arbitrage, we must define some concepts and variables. 

\subsection{Preliminaries}
\label{subsec:model-preliminaries}

We begin by assuming that an arbitrageur identifies an opportunity to arbitrage the pools of the $X\text{-}Y$ token pair in two different rollups, $A$ and $B$. 
The arbitrage opportunity is identified at the end of the last sequenced block of each rollup. At this time, the $X\text{-}Y$ pool in rollup $A$ has a price of $P_A$ and token reserves of $(x_A, y_A)$, while the same pool in rollup $B$ has a price of $P_B$ and token reserves of $(x_B, y_B)$. Note that we are considering prices denominated in token $Y$. In other words, $P_A = \nicefrac{y_A}{x_A}$ and $P_B = \nicefrac{y_B}{x_B}$. Without loss of generality, let's assume that $P_A > P_B$.

We further assume that the arbitrageur maintains liquidity on both rollups, which is kept in the target tokens $X$ and $Y$. Concretely, the arbitrageur's liquidity is $L^X = L_A^X + L_B^X$ (for token $X$) and $L^Y = L_A^Y + L_B^Y$ (for token $Y$). We can value the total liquidity of the arbitrageur in units of token $Y$ using an external price $P_\text{ext}$, and thus, $L = L^Y + L^X \cdot P_\text{ext}$. Note that $P_\text{ext}$ is a theoretical price representing how the arbitrageur values its liquidity. It can be thought of as the price coming from an external source (e.g. an exchange or an oracle) or the price the arbitrage experiences when they settle their liquidity in a future time.

In practice, arbitrageurs do not maintain their liquidity in different tokens, preferring to hedge currency risk by only holding stable tokens such as USDC. In our case, this would mean maintaining liquidity in the stable token and converting back and forth between the target tokens ($X$ and $Y$) and the stable token. However, looking at the impact on liquidity for the target tokens allows for a simpler model while distilling the key aspects of how atomicity impacts arbitrage profit extraction across different scenarios.

In this setup, the arbitrageur will perform two swaps (one in each rollup) to extract this arbitrage opportunity:
\begin{itemize}
    \item Swap $S_B$ in rollup $B$: Pay $\Delta y_B$ units of token $Y$ and receive $\Delta x_B$ units of token $X$ in rollup $B$.
    \item Swap $S_A$ in rollup $A$: Pay $\Delta x_A$ units of token $X$ and receive $\Delta y_A$ units of token $Y$.
\end{itemize}
We define $\mathcal{F}_{S_A}$ and $\mathcal{F}_{S_B}$ as the random variables representing whether the swaps $S_A$ and $S_B$ (respectively) fail. These variables take the value 1 if the swap fails and 0 if the swap is successful. Here, we assume that the failure probabilities for each rollup are independent.

It is important to note that we are ignoring transaction costs in our model. This assumption allows us to avoid converting these costs (usually denominated in ETH) to the target token $Y$, simplifying the analysis. On the other hand, given the current state of rollups, we expect transaction costs to be low enough not to change the conclusions substantially. 

\subsection{Profit Variables}
\label{subsec:model-profit-vars}

Given the arbitrage opportunity defined above, the profit an arbitrageur will extract is simply the difference in liquidity resulting from the swaps $S_A$ and $S_B$, which ultimately depends on the trade sizes ($\Delta x_A$, $\Delta x_B$, $\Delta y_A$, and $\Delta y_B$) and the failure outcomes ($\mathcal{F}_{S_A}$ and $\mathcal{F}_{S_B}$).

Under the non-atomic sequencing regime, one swap can fail while the other does not. Therefore, the difference in liquidity for each target token after the arbitrage is defined as:

\begin{equation}
    \Delta L^X_{\text{non-atomic}} = \Delta x_B \cdot (1-\mathcal{F}_{S_B}) - \Delta x_A \cdot (1-\mathcal{F}_{S_A})
\end{equation}
and

\begin{equation}
    \Delta L^Y_{\text{non-atomic}} = \Delta y_A \cdot (1-\mathcal{F}_{S_A}) - \Delta y_B \cdot (1-\mathcal{F}_{S_B})
\end{equation}
On the other hand, under the atomic execution regime, if one of the swaps reverts, the other swap will also revert. Thus, under this regime:

\begin{equation}
    \Delta L^X_{\text{atomic}} =
        \begin{cases}
          \Delta x_B - \Delta x_A & \text{if} \quad \mathcal{F}_{S_A}=\mathcal{F}_{S_B}=0\\
          0 & \text{otherwise}
        \end{cases}
\end{equation}
and 
\begin{equation}
    \Delta L^Y_{\text{atomic}} =
        \begin{cases}
          \Delta y_A - \Delta y_B & \text{if} \quad \mathcal{F}_{S_A}=\mathcal{F}_{S_B}=0\\
          0 & \text{otherwise}
        \end{cases}
\end{equation}
Using the external price $P_\text{ext}$, we can define the overall profit under each sequencing regime as:

\begin{equation}
    \text{Profit}_{i \in \{\text{non-atomic},  \text{atomic}\}} = \Delta L^Y_i + \Delta L^X_i \cdot P_\text{ext}
\end{equation}
Now, we only need to derive the optimal trade sizes $\Delta x_A$, $\Delta x_B$, $\Delta y_A$, and $\Delta y_B$, which we do in the following subsection.

\subsection{Trade Sizes}
\label{subsec:model-trade-sizes}

To derive the trade sizes of the two swaps, we assume that both pools are \glspl{CPMM}, charge the same trading fee $f$, and that the arbitrageur will execute the optimal trade (i.e., sizing their trade to extract the maximal value from the two target pools). We will further assume that, from the time the arbitrage opportunity is identified to the time the arbitrage trade is executed, no uninformed traders will submit further transactions that shift the prices in affected pools.

In the swap $S_B$, the arbitrageur sends $\Delta y_B$ units of token $Y$ to the pool and pays a fee of $f$. Here, we consider that the \gls{DEX} is processing fees outside of the pool reserves, which means that when a trader wishes to swap a given amount of tokens and pay $\Delta y$, only part of this payment goes to the pool reserve. Concretely, $(1-f) \cdot \Delta y$ is added to the pool reserves, while $f\cdot \Delta y$ is paid to \glspl{LP}. This is the case of Uniswap V3 pools, for instance.

We can derive how many $X$ tokens the arbitrageur receives in a trade of $\Delta y_B$ by using the property of \gls{CPMM} pools in which the product of the two token reserves must always be a constant:
\begin{equation}
\label{eq:delta-x-b}
    x_B \cdot y_B = \left[x_B - \Delta x_B\right]\left[y_B + (1-f)\Delta y_B\right]  \Longleftrightarrow \Delta x_B = \frac{x_B(1-f)\Delta y_B}{y_B + (1-f)\Delta y_B}
\end{equation}
If the arbitrageur executes this trade, the price after the trade will be:

\begin{equation}
    P_B^\text{end} = 
    \frac{y_B +(1-f)\Delta y_B}{x_B - \Delta x_B}= \frac{y_B +(1-f)\Delta y_B}{x_B - \frac{x_B(1-f)\Delta y_B}{y_B + (1-f)\Delta y_B}}= 
    \frac{\left[ y_B +(1-f)\Delta y_B \right]^2}{x_B \cdot y_B}
\end{equation}

%
Using similar logic for the swap $S_A$, the arbitrageur pays $\Delta x_A$ units of token $X$ and receives the following units of token $Y$:

\begin{equation}
\label{eq:delta-y-a}
    \Delta y_A = \frac{y_A(1-f)\Delta x_A}{x_A + (1-f)\Delta x_A}
\end{equation}

And, at the end of the trade, the price of the pool will be:

\begin{equation}
    P_A^\text{end} = \frac{x_A \cdot y_A}{\left[ x_A + (1-f)\Delta x_A \right]^2}
\end{equation}
When the arbitrageur executes the optimal trade, they will pay in rollup $A$ the same units of $X$ tokens they received in rollup $B$, which means that $\Delta x_A = \Delta x_B$. With this equality and equations \ref{eq:delta-x-b} and \ref{eq:delta-y-a}, we can describe the trade sizes $\Delta x_A$, $\Delta x_B$, and $\Delta y_A$ based on the optimal initial size $\Delta y_B$. 

As for $\Delta y_B$, the optimal trade occurs when the prices (excluding fees) in both pools at the end of the trade are equal, i.e., $(1-f)^2P_A^\text{end} = P_B^\text{end}$. We can use this to derive $\Delta y_B$:

\begin{equation}
    (1-f)^2P_A^\text{end} = P_B^\text{end} \Longleftrightarrow \Delta y_B = 
        \frac{(1-f)\sqrt{x_A \cdot y_A \cdot x_B \cdot y_B} - x_A \cdot y_B}{(1-f)x_A + (1-f)^2x_B}
\end{equation}

%
%
%
%
%
%



\section{Atomicity Profit Conditions}
\label{subsec:profit-conditions}

Based on the model developed in Section~\ref{sec:model}, the impact on arbitrage profits of moving from a non-atomic regime to an atomic regime depends on the combined outcome of the random variables $\mathcal{F}_{S_A}$ and $\mathcal{F}_{S_B}$, which represent whether each swap fails. Recall that they take the value~1 if the swap fails and~0 otherwise.

There are four possible combined outcomes for these two variables. For each, we can describe the difference in arbitrage profits between the atomic and non-atomic regimes (i.e., $\text{Profit}_\text{diff}:=\text{Profit}_\text{atomic}-\text{Profit}_\text{non-atomic}$):

\vspace{-2ex}

\begin{itemize}
    \item $\mathcal{F}_{S_A} = 0 \cap \mathcal{F}_{S_B} = 0$. In this outcome, both swaps execute, and thus, the difference in arbitrage profits between the two regimes is zero.
    \item $\mathcal{F}_{S_A} = 1 \cap \mathcal{F}_{S_B} = 1$. In this outcome, both swaps fail. Thus, the difference in arbitrage profits between the two regimes is again zero.
    \item $\mathcal{F}_{S_A} = 1 \cap \mathcal{F}_{S_B} = 0$. In this outcome, swap $S_A$ fails, but swap $S_B$ executes. Here, there is a difference since, in the atomic regime, both swaps would be reverted. Therefore, $\text{Profit}_\text{diff} = 0 - (\Delta x_B P_\text{ext} - \Delta y_B) = \Delta y_B - \Delta x_B P_\text{ext}$
    \item $\mathcal{F}_{S_A} = 0 \cap \mathcal{F}_{S_B} = 1$. In this outcome, swap $S_A$ executes, while swap $S_B$ fails. Again, there is a difference in this combined outcome since, in the atomic regime, both swaps would be reverted. Therefore, $\text{Profit}_\text{diff} = 0 - (\Delta y_A - \Delta x_A P_\text{ext}) = \Delta x_A P_\text{ext} - \Delta y_A$
\end{itemize}
Now, if we define $f_A$ and $f_B$ as the probability of swaps $S_A$ and $S_B$ failing, respectively, we can describe the expected value of the profit difference as follows:

\vspace{-3.5ex}

\begin{align}
    &\nonumber \mathbb{E}[\text{Profit}_\text{diff}] = \\
        &\nonumber =(\Delta y_B - \Delta x_B P_\text{ext}) \cdot P[\mathcal{F}_{S_A} = 1 \cap \mathcal{F}_{S_B} = 0] +\\
        &\nonumber\quad \quad (\Delta x_A P_\text{ext} - \Delta y_A) \cdot P[\mathcal{F}_{S_A} = 0 \cap \mathcal{F}_{S_B} = 1]\\
        &\nonumber = (\Delta y_B - \Delta x_B P_\text{ext}) \cdot f_A \cdot (1-f_B) +  (\Delta x_B P_\text{ext} - \Delta y_A) \cdot (1-f_A) \cdot f_B\\
        &= f_A(\Delta y_B - \Delta x_B P_\text{ext}) + f_B(\Delta x_A P_\text{ext} - \Delta y_A) + f_Af_B(\Delta y_A - \Delta y_B)
\label{eq:profit-diff}
\end{align}
%
%
Interestingly, we can rewrite equation \ref{eq:profit-diff} in terms of the price paid by the arbitrageur in each swap, namely, $P^*_A = \nicefrac{\Delta y_A}{\Delta x_A}$ and $P^*_B = \nicefrac{\Delta y_B}{\Delta x_B}$. Note that here we are using again the fact that $\Delta x_A = \Delta x_B$:

\begin{equation}
    \mathbb{E}[\text{Profit}_\text{diff}] = \Delta x_B\big[ f_A(P^*_B - P_\text{ext}) + f_B(P_\text{ext} - P^*_A) + f_Af_B(P^*_A - P^*_B)\big]
\label{eq:profit-diff-v2}
\end{equation}
%
%
%
%
Equation \ref{eq:profit-diff-v2} highlights that the expected gain that an arbitrageur will experience when switching from a non-atomic regime to an atomic regime ultimately depends on a few key parameters.

\begin{figure}[h]
\centering
\captionsetup{justification=centering}
\begin{subfigure}{.35\textwidth}
  \centering
  \includegraphics[width=\linewidth]{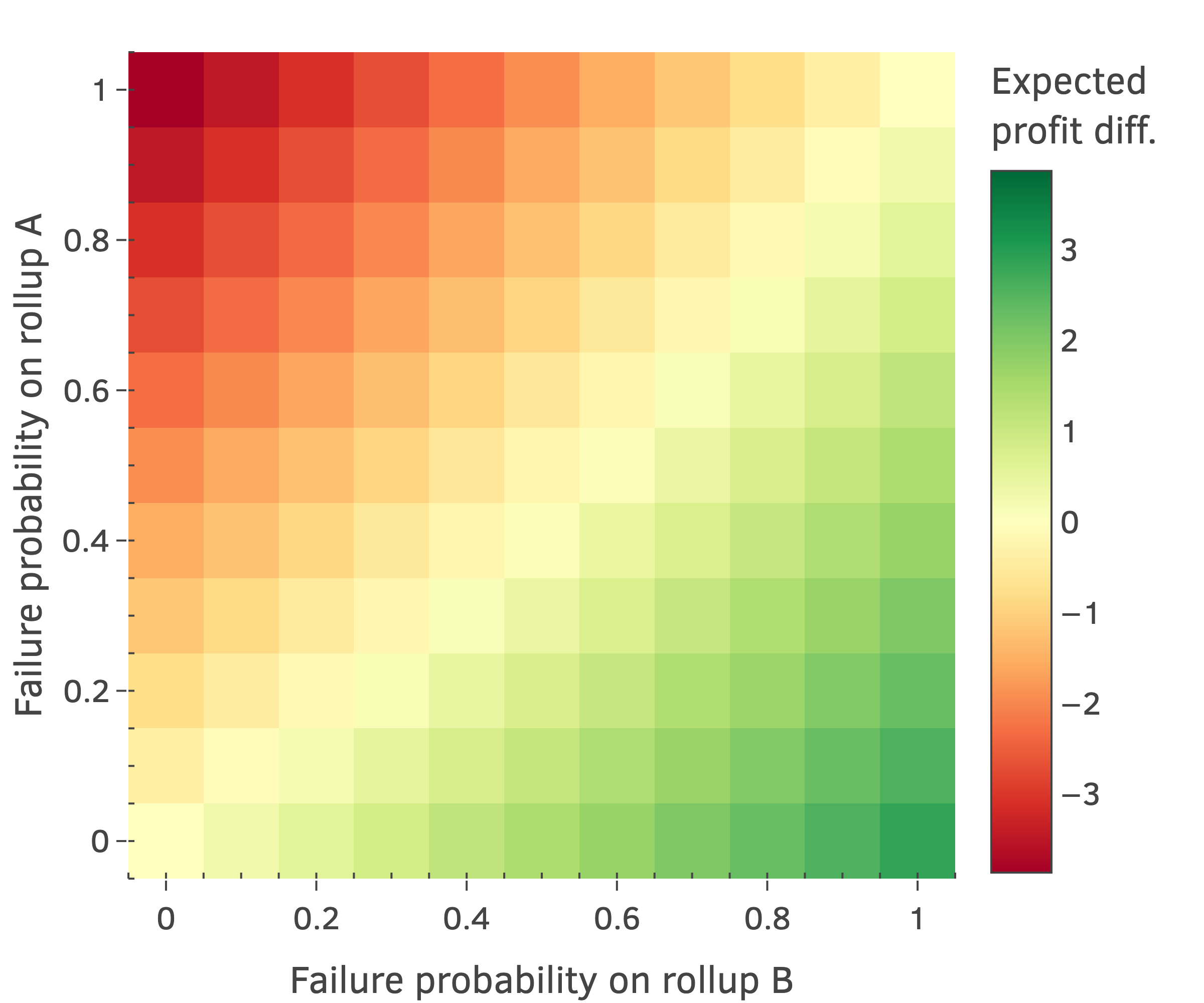}
  \caption{$P^*_B < P^*_A < P_\text{ext}$\\ ($P_\text{ext}\approx1.020$)}
  \label{fig:high-price}
\end{subfigure}%
\begin{subfigure}{.35\textwidth}
  \centering
  \includegraphics[width=\linewidth]{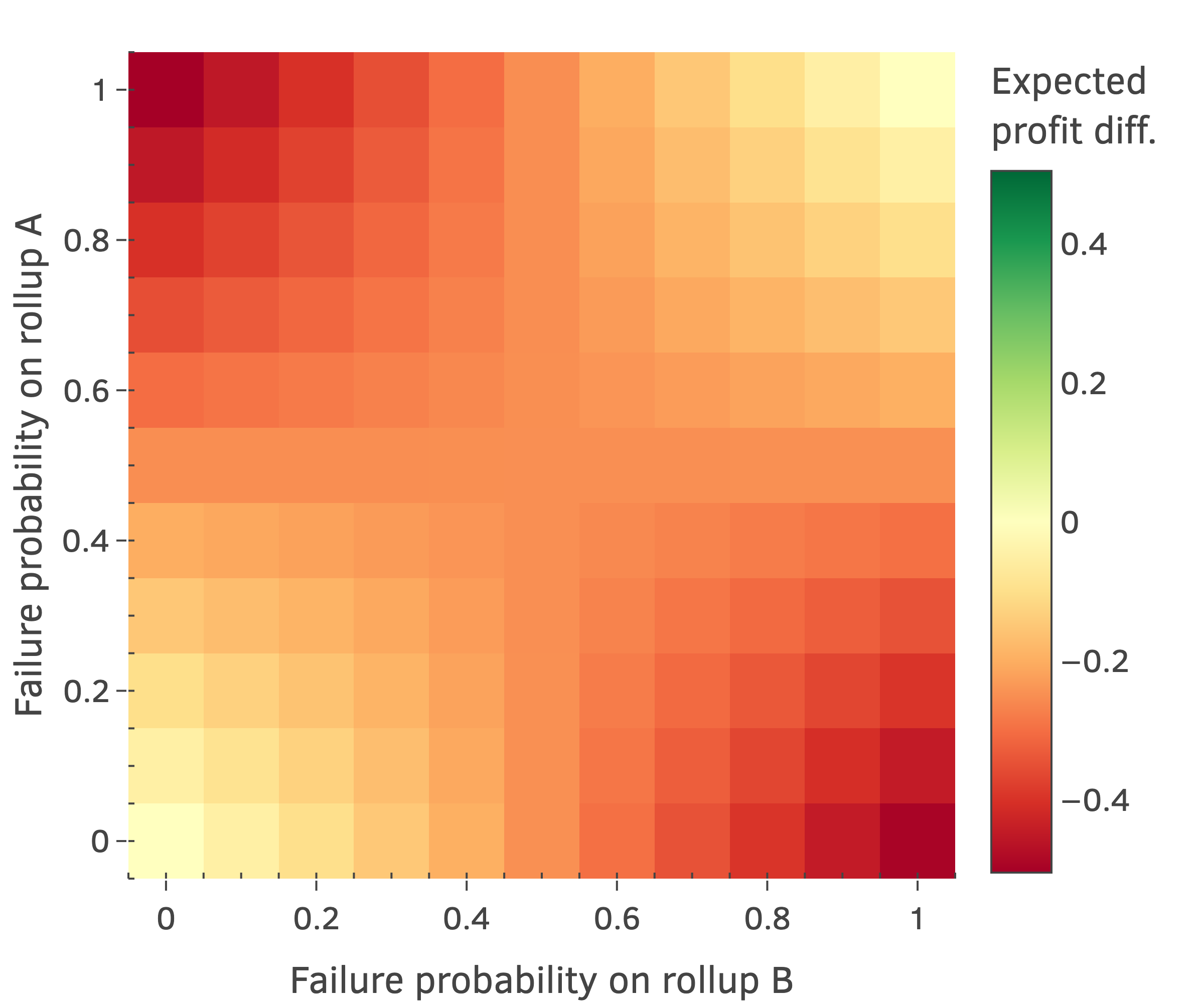}
  \caption{$P^*_B < P_\text{ext} < P^*_A$\\ ($P_\text{ext}=1.005$)}
  \label{fig:middle-price}
\end{subfigure}%
\begin{subfigure}{.35\textwidth}
  \centering
  \includegraphics[width=\linewidth]{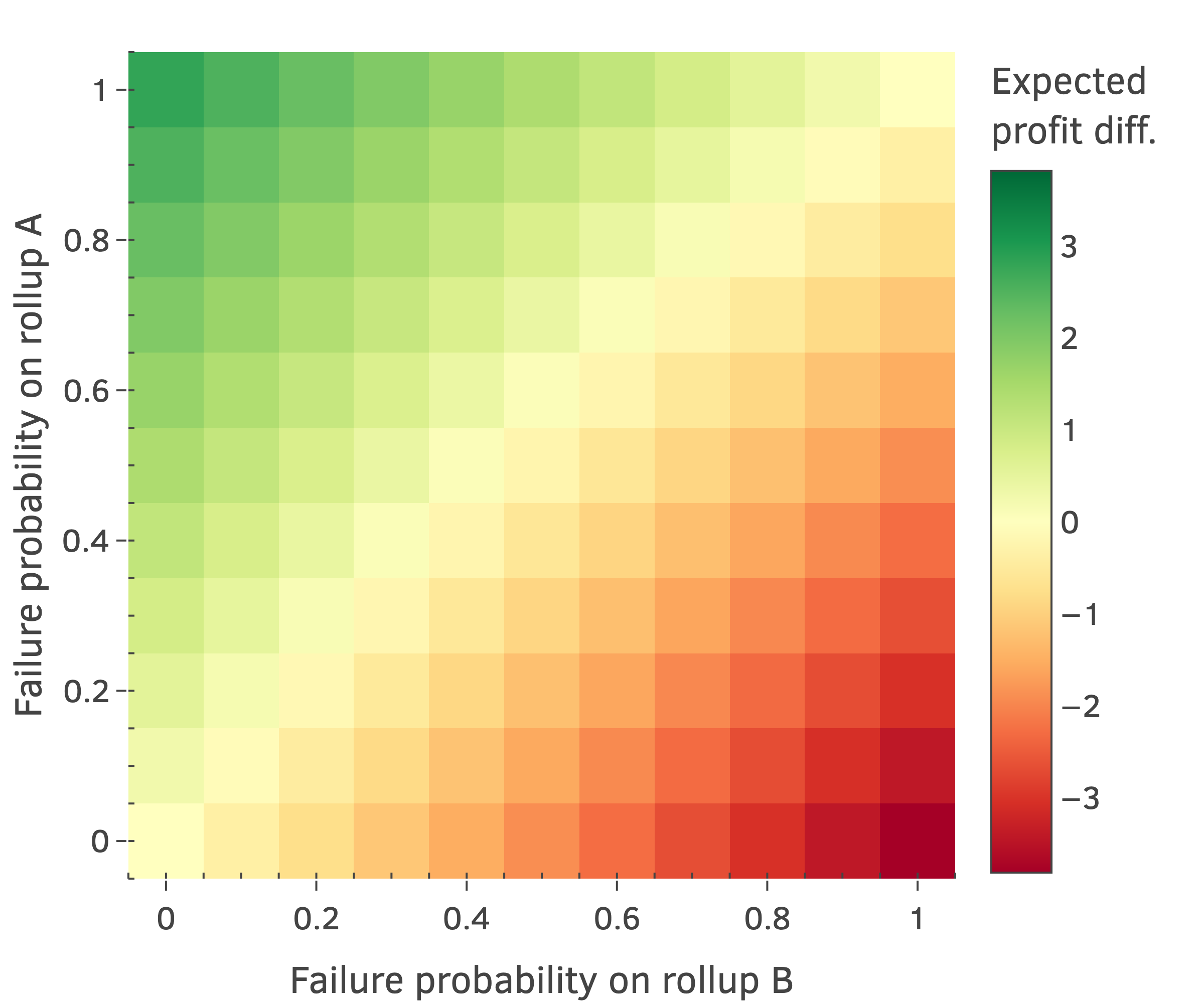}
  \caption{$P_\text{ext} < P^*_B < P^*_A$\\ ($P_\text{ext}\approx0.990$)}
  \label{fig:low-price}
\end{subfigure}
\caption{Expected value of the difference in arbitrage profits between the atomic and non-atomic regimes for varying failure probabilities and relative external prices. Pool states are kept unchanged: $f=0.05\%$, $P_A=1.01$, $P_B=1$, and $y_A=y_B=100,000$}
\label{fig:param-heat-maps}
\end{figure}

\begin{enumerate}
\item We have the trade size $\Delta x_B$. Recall from Section~\ref{sec:model} that the trade size is determined by the state of the pools in each rollup, namely, the token reserves, the price difference, and the trading fee. In general, the larger the pools' reserves and the price difference, the larger the optimal trade sizes. Since $\Delta x_B >0$, the pools' state does not control whether the difference is negative or positive on average. Instead, it has a multiplicative effect on the expected profit difference, controlling the size of this difference.

\item We have the external price $P_\text{ext}$ and its relative position to the prices experienced by the arbitrageur in their optimal trade, $P^*_A$ and $P^*_B$. These prices are always between the initial price of the pool before the arbitrage and the end price after the arbitrage is executed (i.e., $P_A > P^*_A > P^\text{end}_A$ and $P_B < P^*_B < P^\text{end}_B$). 
\item  There are the failure probabilities $f_A$ and $f_B$, which we will analyze together with the external price. Figure \ref{fig:param-heat-maps} provides an example for each of the three possible configurations of the relative position of $P_\text{ext}$ and $P^*_A$ and $P^*_B$, and the full range of failure probabilities $f_A$ and $f_B$. 

\end{enumerate}

We should highlight that this formula does not depend on the exact design of the \gls{AMM}. To ease the implementation of the entire simulation, we assumed it was a \gls{CPMM} as it made the derivation of the optimal trade sizes simpler. However, this formula would also hold for more complex designs such as Uniswap v3.

When the external price is \emph{larger} than both pool prices (Figure \ref{fig:high-price}), the expected profit difference can be positive or negative, depending on the failure probabilities. If failures are more likely on rollup $B$, the difference is positive, meaning that the arbitrageur will profit on average by switching to the atomic regime. However, if failures are more likely on rollup $A$, the difference is negative, and switching is no longer profitable. 

Intuitively, this relationship makes sense. When the external price has a larger difference to the price on rollup $B$ than the price differences in the two rollups, it would be better to only execute the swap on rollup $B$ than to arbitrage it against rollup~$A$. Therefore, having swap $S_A$ failing and swap $S_B$ executing leads to a net gain for the arbitrageur when valued against this external price (which means that atomicity is worse for the arbitrageur).

On the other hand, when the external price is \emph{smaller} than both pool prices (Figure \ref{fig:low-price}), the relationship is inverted. In this case, the rationale is similar, and switching to an atomic regime is only advantageous when failures are more likely on rollup $A$ since simply swaping on rollup $A$ generates more profit than arbitraging it against rollup $B$.


Finally, there is the case where the external price is between the two pool prices (Figure~\ref{fig:middle-price}). Interestingly, the expected profit difference is always negative in this case, independently of the failure probabilities. Similarly to the previous cases, when we value liquidity using an external price, and one of the swaps fails and the other executes, we are, in a way, arbitraging the pool that did not fail against the external price. When the external is between the prices in each pool, having only one swap failing is always better than having both reverting, as we would collect some additional profit from arbitraging the pool that did not fail against the external price.

Focusing on the failing probabilities, there is a special case we can analyze. If the two failure probabilities are equal (i.e., $f_A=f_B$), the expected profit difference is always negative, meaning that, on average, the arbitrageur should \emph{stick with the status quo} without atomicity. This result comes directly from Equation~\ref{eq:profit-diff-v2}:

\vspace{-3ex}

\begin{align}
    &\nonumber \mathbb{E}[\text{Profit}_\text{diff}] = \\
    &\nonumber = \Delta x_B\big[ f_A(P^*_B - P_\text{ext}) + f_A(P_\text{ext} - P^*_A) + f_A^2(P^*_A - P^*_B)\big] =\\
    &\nonumber = \Delta x_B\big[ f_A(P^*_B - P^*_A) + f_A^2(P^*_A - P^*_B)\big] =\\
    &= \Delta x_B \cdot f_A \cdot (1-f_A) \cdot (P^*_B - P^*_A) < 0
\label{eq:profit-diff-cond}
\end{align}

\section{Related Work}
\label{sec:related-work}

The seminal work from \citet{daian_flash_2020} laid the groundwork for understanding \gls{MEV} by demonstrating how block producers could exploit transaction ordering on Ethereum to capture arbitrage and frontrunning profits. Since then, multiple empirical studies have examined \gls{MEV} on Ethereum by uncovering common strategies and measuring their prevalence and impact \cite{torres_frontrunner_2021, qin_quantifying_2022}.

Recent work has extended the analysis into the Layer-2 domain. For instance, \citet{torres_rolling_2024, ha_flashbabies_2021} and \citet{bagourd_quantifying_2023} present different measurements of MEV across popular rollups and other Layer-2s, revealing the most common strategies, their volume, and the corresponding profit. Complementarily, \citet{gogol_layer-2_2024} examines arbitrage across different rollups and identifies many untapped arbitrage opportunities resulting from the non-atomic nature of cross-rollup transactions. Finally, \citet{oz_pandoras_2025} systematically analyzes non-atomic cross-chain arbitrage strategies across multiple L2s. It reveals that liquidity fragmentation across heterogeneous blockchain networks creates substantial arbitrage opportunities while also highlighting the inherent challenges posed by non-atomic execution. 

Beyond empirical analyses, \citet{mcmenamin_sok_2023} categorizes MEV extraction methods across multiple domains and outlines proposals for shared sequencers coordinating transaction ordering across chains. In addition, \citet{mamageishvili_shared_2023} models the economic incentives under different sequencing regimes. Their results suggest that while a unified sequencer could facilitate atomic cross-chain arbitrage, it may also intensify latency competition and not necessarily increase overall sequencer revenue compared to independent rollup-specific sequencers.
\section{Conclusions}
\label{sec:conclusion}

Our work studies cross-rollup arbitrage in the context of a shared sequencing system offering atomic execution, a feature that ensures either all or none of a sequence of arbitrage transactions across multiple rollups are executed. Here, we investigate whether atomic execution is sufficient to significantly boost arbitrage profits and, in turn, sequencer revenue.

Our results reveal that arbitrage profits do not always improve under atomic execution, and thus, this feature alone is not enough to convince both arbitrageurs and rollup operators to switch to this new approach. When considering a case where an arbitrageur exploits an opportunity between two \gls{CPMM} pools and values the final profit using an external price, we find that whether switching from non-atomic to atomic execution is net positive for the arbitrageur depends on the failure probabilities of the swaps in each rollup and the relative difference of the external price to the pool prices.

This work could be extended in multiple ways. Firstly, we assume that the arbitrageur maintains their liquidity in the tokens being arbitraged. However, arbitrageurs may keep liquidity in a stable token and convert it on demand to address volatility; our model could be extended to account for this conversion. Secondly, we do not consider transaction costs. Although it is currently low and likely to remain such for rollups, adding this cost would be another possible extension to the model. Thirdly, and more importantly, one could explore how prevalent the scenarios in which atomic execution is not beneficial to an arbitrageur are. This would require a detailed empirical analysis of various pools across different deployed rollups and varying time periods.

\bibliographystyle{splncs04nat}
\bibliography{references}

\begin{thebibliography}{15}
\providecommand{\natexlab}[1]{#1}
\providecommand{\url}[1]{\texttt{#1}}
\providecommand{\urlprefix}{URL }
\expandafter\ifx\csname urlstyle\endcsname\relax
  \providecommand{\doi}[1]{doi:\discretionary{}{}{}#1}\else
  \providecommand{\doi}{doi:\discretionary{}{}{}\begingroup \urlstyle{rm}\Url}\fi

\bibitem[{noa(2024)}]{noauthor_astria_2024}
Astria: {The} {Shared} {Sequencer} {Network} (Feb 2024), \urlprefix\url{https://www.astria.org/blog/astria-the-shared-sequencer-network}

\bibitem[{Bagourd and Francois(2023)}]{bagourd_quantifying_2023}
Bagourd, A., Francois, L.G.: Quantifying {MEV} {On} {Layer} 2 {Networks} (Aug 2023), \doi{10.48550/arXiv.2309.00629}, \urlprefix\url{http://arxiv.org/abs/2309.00629}, arXiv:2309.00629 [cs, q-fin]

\bibitem[{Daian et~al.(2020)Daian, Goldfeder, Kell, Li, Zhao, Bentov, Breidenbach, and Juels}]{daian_flash_2020}
Daian, P., Goldfeder, S., Kell, T., Li, Y., Zhao, X., Bentov, I., Breidenbach, L., Juels, A.: Flash {Boys} 2.0: {Frontrunning} in {Decentralized} {Exchanges}, {Miner} {Extractable} {Value}, and {Consensus} {Instability}. In: 2020 {IEEE} {Symposium} on {Security} and {Privacy} ({SP}), pp. 910--927 (May 2020), \doi{10.1109/SP40000.2020.00040}, \urlprefix\url{https://ieeexplore.ieee.org/abstract/document/9152675}, iSSN: 2375-1207

\bibitem[{Gogol et~al.(2024)Gogol, Messias, Miori, Tessone, and Livshits}]{gogol_layer-2_2024}
Gogol, K., Messias, J., Miori, D., Tessone, C., Livshits, B.: Layer-2 {Arbitrage}: {An} {Empirical} {Analysis} of {Swap} {Dynamics} and {Price} {Disparities} on {Rollups} (Jun 2024), \doi{10.48550/arXiv.2406.02172}, \urlprefix\url{http://arxiv.org/abs/2406.02172}, arXiv:2406.02172 [cs]

\bibitem[{{Ha} et~al.(2021){Ha}, {Vlachou}, {Kilbourn}, and {De Michellis}}]{ha_flashbabies_2021}
{Ha}, {Vlachou}, {Kilbourn}, {De Michellis}: {FlashBabies} - {MEV} on {L2} (Dec 2021)

\bibitem[{Mamageishvili and Schlegel(2023)}]{mamageishvili_shared_2023}
Mamageishvili, A., Schlegel, J.C.: Shared {Sequencing} and {Latency} {Competition} as a {Noisy} {Contest} (Oct 2023), \urlprefix\url{http://arxiv.org/abs/2310.02390}, arXiv:2310.02390 [cs]

\bibitem[{McMenamin(2023)}]{mcmenamin_sok_2023}
McMenamin, C.: {SoK}: {Cross}-{Domain} {MEV} (Aug 2023), \urlprefix\url{http://arxiv.org/abs/2308.04159}, arXiv:2308.04159 [cs]

\bibitem[{Motepalli et~al.(2023)Motepalli, Freitas, and Livshits}]{motepalli_sok_2023}
Motepalli, S., Freitas, L., Livshits, B.: {SoK}: {Decentralized} {Sequencers} for {Rollups} (Oct 2023), \doi{10.48550/arXiv.2310.03616}, \urlprefix\url{http://arxiv.org/abs/2310.03616}, arXiv:2310.03616 [cs]

\bibitem[{NodeKit(2024)}]{nodekit_composable_2024}
NodeKit: The {Composable} {Network}: {Unifying} {Chains} with {Javelin}, the first {Superbuilder} (Aug 2024), \urlprefix\url{https://nodekit-tinywins.vercel.app/posts/the-composable-network}

\bibitem[{Qin et~al.(2022)Qin, Zhou, and Gervais}]{qin_quantifying_2022}
Qin, K., Zhou, L., Gervais, A.: Quantifying {Blockchain} {Extractable} {Value}: {How} dark is the forest? In: 2022 {IEEE} {Symposium} on {Security} and {Privacy} ({SP}), pp. 198--214 (May 2022), \doi{10.1109/SP46214.2022.9833734}, \urlprefix\url{https://ieeexplore.ieee.org/abstract/document/9833734}, iSSN: 2375-1207

\bibitem[{Radius(2023)}]{radius_radius_2023}
Radius: Radius: {Trustless} {Shared} {Sequencing} {Layer} (May 2023), \urlprefix\url{https://medium.com/@radius_xyz/radius-trustless-shared-sequencing-layer-b293dfa75db}

\bibitem[{Systems(2023)}]{espresso_systems_espresso_2023}
Systems, E.: The {Espresso} {Sequencer} (Mar 2023), \urlprefix\url{https://hackmd.io/@EspressoSystems/EspressoSequencer}

\bibitem[{Torres et~al.(2021)Torres, Camino, and State}]{torres_frontrunner_2021}
Torres, C.F., Camino, R., State, R.: Frontrunner {Jones} and the {Raiders} of the {Dark} {Forest}: {An} {Empirical} {Study} of {Frontrunning} on the {Ethereum} {Blockchain}. pp. 1343--1359 (2021), ISBN 978-1-939133-24-3, \urlprefix\url{https://www.usenix.org/conference/usenixsecurity21/presentation/torres}

\bibitem[{Torres et~al.(2024)Torres, Mamuti, Weintraub, Nita-Rotaru, and Shinde}]{torres_rolling_2024}
Torres, C.F., Mamuti, A., Weintraub, B., Nita-Rotaru, C., Shinde, S.: Rolling in the {Shadows}: {Analyzing} the {Extraction} of {MEV} {Across} {Layer}-2 {Rollups} (Apr 2024), \urlprefix\url{http://arxiv.org/abs/2405.00138}, arXiv:2405.00138 [cs]

\bibitem[{Öz et~al.(2025)Öz, Torres, Gebele, Rezabek, Mazorra, and Matthes}]{oz_pandoras_2025}
Öz, B., Torres, C.F., Gebele, J., Rezabek, F., Mazorra, B., Matthes, F.: Pandora's {Box}: {Cross}-{Chain} {Arbitrages} in the {Realm} of {Blockchain} {Interoperability} (Jan 2025), \doi{10.48550/arXiv.2501.17335}, \urlprefix\url{http://arxiv.org/abs/2501.17335}, arXiv:2501.17335

\end{thebibliography}


\end{document}